\documentclass[fleqn,12pt]{wlscirep}
\usepackage[utf8]{inputenc}
\usepackage[T1]{fontenc}
\title{Engineering snags for spatial curvature in weaves: Fabrication, mechanics, and inverse design}

\usepackage{caption}
\usepackage{helvet}  
\makeatletter
\renewcommand{\@makecaption}[2]{%
  {\fontfamily{phv}\selectfont\footnotesize\textbf{#1.} #2\par}%
}
\makeatother 

\author[1]{Guowei Wayne Tu}
\author[1,2,*]{Evgueni T. Filipov}
\affil[1]{Deployable and Reconfigurable Structures Laboratory, Department of Civil and Environmental Engineering, University of Michigan, Ann Arbor, MI, USA}
\affil[2]{Deployable and Reconfigurable Structures Laboratory, Department of Mechanical Engineering, University of Michigan, Ann Arbor, MI, USA}
\affil[*]{E-mail: filipov@umich.edu}

\begin{abstract}
Weaving as an old craft has extensive applications in modern science and technology such as smart textiles and intelligent soft robots. However, weaving irregular curved surfaces has been difficult, with prior alternatives requiring curved ribbons and triaxial weaving patterns. In this work, we present a simple strategy to achieve complex spatial curvature by purposely introducing `snags’, a traditionally unwanted textile defect, into dense plain weaves consisting of straight ribbons assembled in a straightforward biaxial network. We detail the fabrication methodology where we pull out ribbons of initially smooth two- (2D) and three-dimensional (3D) plain weaves to form local snags. We show that these local defects cause global curvatures through the propagation of geometric frustration. We then use a reduced-order bar \& hinge model to simulate the mechanics-guided deformation of snagged plain weaves, and we investigate how the curvature scales with system parameters such as the thickness and Young’s modulus of the ribbons. Finally, we introduce an inverse design platform where an evolutionary algorithm is used to inversely compute the optimal snag patterns of smooth plain weaves to approximate arbitrary target surfaces including 2D and 3D woven exoskeletons that fit human legs and elbows, respectively. Engineering snags in plain weaves as a general strategy can pave the way for future design of customizable wearable devices, adaptive soft robots, reconfigurable architecture, and more. 
\end{abstract}
\begin{document}

\flushbottom
\maketitle
%
%
\thispagestyle{empty}

\section{Introduction}\label{Sec:Intro}
As an efficient strategy to assemble one-dimensional (1D) materials into two- (2D) and three-dimensional (3D) structural systems, weaving techniques are being explored in modern science and engineering to make artificial muscles \cite{maziz2017knitting,haines2016new,shin2024woven}, impact-resistant armor \cite{miao2025flexible,wang2017smart,qiu2023impact}, architected metamaterials \cite{surjadi2025double,wang2023molecularly}, and more. When fabricating these woven systems, there are various weaving patterns that can be used to arrange the 1D materials into more complex geometries. Among all weaving patterns, the plain weave pattern is the most straightforward where straight ribbons/fibers are arranged in an one-over-one-under manner \cite{lan2023woven,zhang2022molecular,yang2024single,chen2025implications}. The simplicity of plain weaves makes them highly compatible with automated weaving machines \cite{zhang2023hierarchical,yan2022single}, adaptable to different materials \cite{wang2025smart,lin2024triboelectric,xiong2021functional}, and to different length scales \cite{lu2024intelligent,he2021scalable,shi2021large,zeng2021hierarchical}. 

However, plain weaves lack a capability of forming complex spatial curvatures, which are necessary for many engineering applications \cite{schamberger2023curvature,buckner2020roboticizing,yang2022gaussian}. To approximate intricate 3D curved surfaces, one common strategy is sparse triaxial weaving with originally curved ribbons \cite{baek2021smooth,ren20213d,poincloux2023indentation}. Weaving non-uniform curved ribbons in a tri-axial manner is time-consuming, non-intuitional, and less adaptive to different materials compared to plain weaves \cite{lewandowska2017triaxial,barman2024textile,bilisik2012multiaxis}. Besides, the sparsity of tri-axially woven surfaces undermines the structural strength and water/ballistic resistance of the system \cite{el2016stab,shi2023effect}. Another common strategy to construct 3D curvatures is mold forming/draping of planar woven sheets \cite{khiem2024modeling,khiem2018averaging,mei2021analysis,chen2025experimental}. These mold forming processes cause irreversible plastic damage to the fibers and weaken the mechanical performance of the woven systems.

\begin{figure}[!htb] 
\centering
\includegraphics[width=17.8cm]{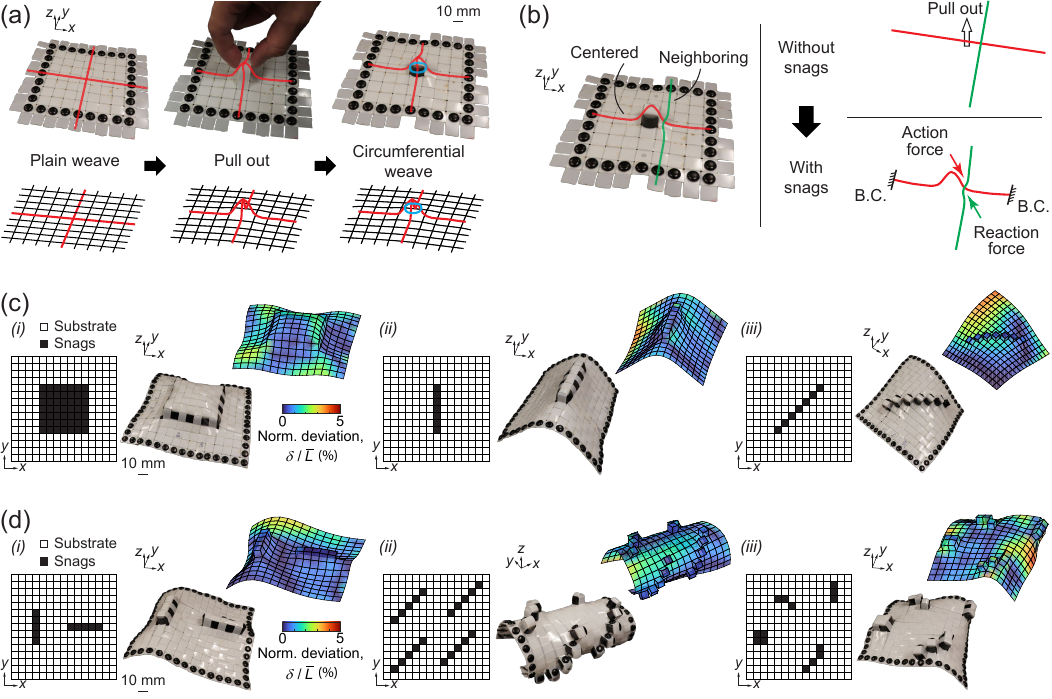}
\caption{\textbf{Engineering snags in plain weaves to achieve spatial curvature}. (a) Fabrication of a snag in a densely plain-woven flat sheet; (b) How a local snag causes curvature through the propagation of action-reaction force pairs; `B.C.’ stands for `boundary condition’; (c) Variations of the size, shape, and direction of the snag in (a) and the resulting geometries; (c, \textit{i}) shows a larger yet still square snag, (c, \textit{ii}) shows a rectangular strip-like snag, and (c, \textit{iii}) shows a diagonal strip-like snag; (d) Combinations of the three types of snags in (c) and the resulting geometries; (d, \textit{i}) shows a combination of two rectangular snags, which results in a skate fish-like shape; (d, \textit{ii}) shows a combination of four diagonal snags, which results in an oblique cylindrical shape; (d, \textit{iii}) shows a random combination of square, rectangular, and diagonal snags, which results in a random complex shape. In each case in (c) and (d), the left shows the 2D snag patterns, the middle shows the physical prototypes, and the right shows the simulated shapes where the colormap represents the normalized deviation between the simulated and scanned geometries $\delta/\overline{L}$ where $\delta$ is the deviation (Hausdorff distances at each node) and $\overline{L}$ is the average length of all ribbons used.}\label{fig:2DSnag}
\end{figure}

In this work, we introduce a weaving technique where we intentionally introduce simple \textit{snags} into dense plain weaves to achieve complex spatial curvature. As a traditionally unwanted defect in woven fabrics where a ribbon/fiber is accidentally pulled out of place and misaligned from its intended interlacing path, snags distort the surface and can cause unexpected deformation and curvature \cite{weerasinghe2022impact,yu2021simulating,nilakantan2015experimental}. Here, we systematically engineer these localized disruptions to alter the geometry of 2D and 3D plain-woven surfaces and structures in a predictable and programmable manner. Some examples are shown in Figs. \ref{fig:2DSnag} and \ref{fig:3DSnag}. 

We first discuss the fabrication of snagged plain weaves where we follow a pre-designed \textit{snag pattern} and introduce those snags into an originally 2D or 3D plain weave in a simple and systematic way (Section \ref{Sec:Snags}). We also explain how these local geometric frustrations cause a global curvature. Then, we provide a mechanics-based reduced-order bar \& hinge model to simulate the geometry of snagged plain weaves (Section \ref{Sec:Modeling}). Using this mechanics model, we numerically and experimentally investigate how spatial curvatures of snagged plain weaves scale with essential design parameters including the size of the surface, the thickness of the ribbons, and the Young's modulus of the material (Section \ref{Sec:Scaling}). After gaining insights into the coupling between the mechanics and geometry, we establish an inverse design framework where we can find an optimal snag pattern once a target surface is given (Section \ref{Sec:InverseDesign}). We give real-life examples where 2D and 3D snagged plain-woven surfaces can fit onto a human leg and elbow, which is a potential solution to customized woven exoskeleton suits \cite{tian2023implant}. As a general simple strategy to program the curvature of densely plain-woven systems, engineering snags can make a paradigm shift in the future design of next-generation wearable electronics, intelligent soft robots, flexible metamaterials, and more. 

\section{Engineering snags in plain weaves}\label{Sec:Snags}
In this section, we detail how we fabricate snags in 2D plain weaves, how local snags enable global curvatures, and how we generalize the snagging strategy to complex 3D plain weaves. 

\subsection{Snags in 2D plain-woven sheets}
Figure \ref{fig:2DSnag}(a) shows how we start the fabrication with a 2D square densely plain-woven sheet where the dimension is 9$\times$9 $\mathrm{unit}^2$ and all constituent ribbons are secured at the boundary by split pins. All physical models shown here are made from Mylar\textsuperscript{\textregistered} polyester that is 7.5 mil (0.1905 mm) thick, and the width of the ribbons is 10 mm. The plain weave is densely packed, while a small gap $d=0.5 \ \mathrm{mm}$ is included between the crossing ribbons such that the length of each ribbon is $(w+d)N_c$ where $w$ is the width of ribbons and $N_c$ is the number of crossing ribbons. Then, we vertically pull out the two ribbons that cross in the center of the sheet while we re-secure both ribbons at their new boundaries. Finally, we use a circumferential ribbon to weave around the pulled-out part. In this way, we introduce a snag to the 2D woven sheet. The snag causes the two centered ribbons to protrude from the sheet, but it also causes the adjacent ribbons to deform vertically along the positive \textit{z} direction and thus leads to curvature of the sheet. A local snag causes curvature because the neighboring ribbons impose an action force downward on the snagged centered ribbons as the centered ribbons move upward, and meanwhile the neighboring ribbons are subject to a reaction force upward (see Fig. \ref{fig:2DSnag}(b)). This local perturbation eventually spreads through all ribbons and gives the woven sheet a dome-like shape (even though the magnitude of the dome is small in Fig. \ref{fig:2DSnag}(b)). 

We can change the geometry of the woven sheet by changing the size, shape, or direction of the snag. Figure \ref{fig:2DSnag}(c) gives three examples, where first we make the dome-like shape deeper through a larger yet still square snag (Fig. \ref{fig:2DSnag}(c, \textit{i})), then we make the sheet bend about only one axis of symmetry through a rectangular strip-like snag (Fig. \ref{fig:2DSnag}(c, \textit{ii})), and last we make the sheet bend diagonally through a diagonal strip-like snag (Fig. \ref{fig:2DSnag}(c, \textit{iii})). Further, by combining the three types of fundamental snags in Fig. \ref{fig:2DSnag}(c), we create more complex snag patterns which result in a skate fish-like shape (Fig. \ref{fig:2DSnag}(d, \textit{i})), an oblique cylindrical shape (Fig. \ref{fig:2DSnag}(d, \textit{ii})), and a random complex shape (Fig. \ref{fig:2DSnag}(d, \textit{iii})). In all binary snag patterns in Fig. \ref{fig:2DSnag}(c) and (d), we merge any snag units that share sides (excluding the diagonal units that only share vertices) into a snag block, and we weave the circumferential ribbon of each snag block as a whole. An excellent match is observed between the scanned geometry of the physical prototypes and our mechanics simulation (which is detailed later in Section \ref{Sec:Modeling}), as shown in the colormaps in Fig. \ref{fig:2DSnag}(c) and (d). 

\subsection{Generalizing the snagging strategy to complex 3D surfaces}
\begin{figure}[!htb] 
\centering
\includegraphics[width=17.8cm]{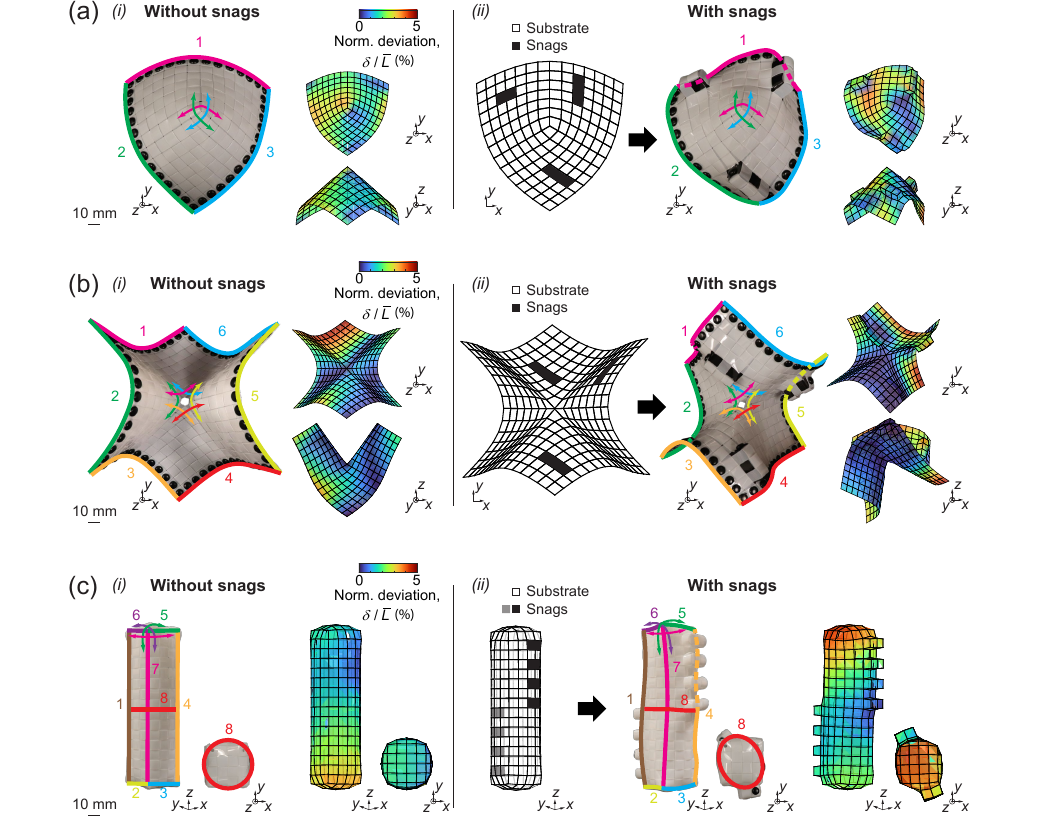}
\caption{\textbf{Generalizing the snagging strategy to 3D plain-woven surfaces}. (a) Comparison between a 3D plain-woven conical surface (\textit{i}) without and (\textit{ii}) with snags; (b) Comparison between a 3D plain-woven saddle surface (\textit{i}) without and (\textit{ii}) with snags; (c) Comparison between a 3D plain-woven column that has a closed surface (\textit{i}) without and (\textit{ii}) with snags. In (a, \textit{i}), (b, \textit{i}), and (c, \textit{i}), the left shows the physical prototypes, and the right shows the simulated shapes. In (a, \textit{ii}), (b, \textit{ii}), and (c, \textit{ii}), the left shows the snag patterns, the middle shows the physical prototypes, and the right shows the simulated shapes. In all simulated shapes in (a), (b), and (c), the colormap represents the normalized deviation between the simulated and scanned geometries $\delta/\overline{L}$ where $\delta$ is the deviation (Hausdorff distances at each node) and $\overline{L}$ is the average length of all ribbons used. In photos of physical prototypes in (a), (b), and (c), the colored arrows indicate the paths of the inter-woven ribbons that meet at the centers (a, b) or the vertices (c) of each 3D surface, while the colored solid lines without arrows highlight the edges of each 3D surface.}\label{fig:3DSnag}
\end{figure}

We use snags to tune the shapes of 2D woven sheets in Fig. \ref{fig:2DSnag}, but we can also generalize this snagging strategy to 3D plain-woven surfaces that have a spatial geometry. Figure \ref{fig:3DSnag}(a, \textit{i}) shows a 3D conical surface made by interweaving twenty-one identical Mylar\textsuperscript{\textregistered} polyester ribbons \cite{tu2025basketweaving}. We then introduce three sparsely distributed snags to this 3D surface to distort the conical shape and break its symmetry (see Fig. \ref{fig:3DSnag}(a, \textit{ii})). This phenomenon of local snags breaking a global symmetry also occurs for a 3D saddle surface made by interweaving forty-two identical Mylar\textsuperscript{\textregistered} polyester ribbons (Fig. \ref{fig:3DSnag}(b)) and a 3D column-like structure with a closed surface composed of a square plain-woven basket and a square cover (Fig. \ref{fig:3DSnag}(c)) \cite{tu2025basketweaving}. The scanned geometry of the physical prototypes match well with our mechanics simulation (which is detailed later in Section \ref{Sec:Modeling}), as shown in the colormaps in Fig. \ref{fig:3DSnag}.

\begin{figure}[!htb] 
\centering
\includegraphics[width=17.8cm]{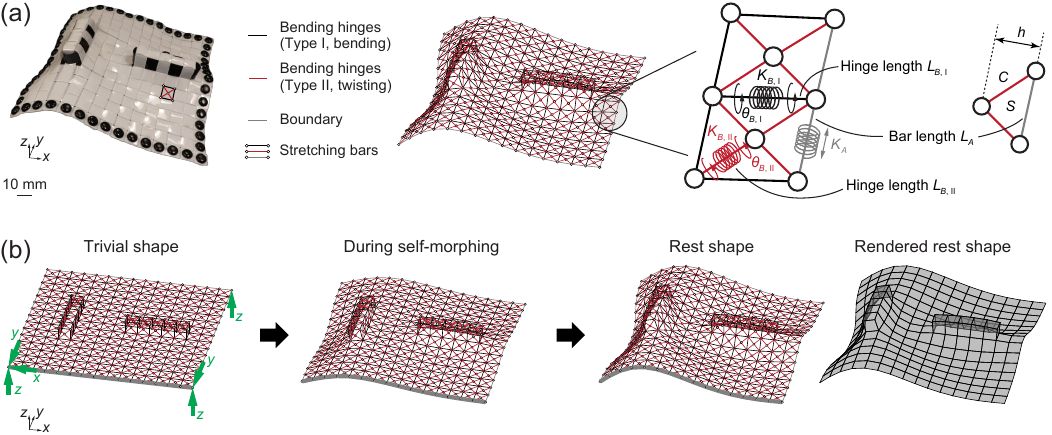}
\caption{\textbf{Bar \& hinge simulation of woven surfaces}. (a) Physical prototype and the corresponding bar \& hinge mesh of a plain-woven sheet with snags. A zoomed in schematic of the different bar and hinge element types is shown on the right. (b) Self-morphing process of the surface in (a) from its trivial shape to its rest shape during bar \& hinge simulation. The green arrows in the trivial shape represent the boundary conditions that eliminate the six rigid body modes of the system during the self-morphing process.}\label{fig:Model}
\end{figure}

\section{Mechanics modeling}\label{Sec:Modeling}
We use a \textit{bar} \& \textit{hinge model} to simulate the geometry of snagged woven surfaces. The bar \& hinge model is a reduced-order finite element method where we use bar elements to characterize stretching and shearing and we use torsional spring (hinge) elements to characterize bending and twisting of a structural system \cite{filipov2017bar,tu2024origami}. For snagged plain-woven surfaces such as the woven sheet in Fig. \ref{fig:Model}(a), the small squares formed by the intersection of woven ribbons become the basic units in the bar \& hinge mesh. The four sides along the outer edge of each square are associated with bending hinges of type I, which characterize out-of-plane ribbon bending along the ribbon; while the four half-diagonals inside each square are associated with bending hinges of type II, which characterize out-of-plane ribbon twisting. All locations of bending hinges also include bar elements which characterize in-plane ribbon stretching and shearing. 

We determine the values of the bar \& hinge parameters, including the bending stiffness $K_{B}$ and stretching stiffness $K_A$, using the material properties of the ribbons and the geometry of the triangular panels in the mesh. The bending stiffness $K_{B}$ is given by
\begin{equation}\label{eqn:KB}
{K_B} = C_B D_{\mathrm{eff}} {\left( {\frac{{{L_B}}}{t_{\mathrm{eff, \ bend}}}} \right)^{1/3}} = C_B \frac{{E{t^3_{\mathrm{eff, \ bend}}}}}{{12(1 - {\nu ^2})}}{\left( {\frac{{{L_B}}}{t_{\mathrm{eff, \ bend}}}} \right)^{1/3}},
\end{equation}
where $C_B$ is a dimensionless constant for bending stiffness calibration \cite{filipov2017bar}, $D_{\mathrm{eff}}=Et_{\mathrm{eff, \ bend}}^{3}/[12(1-\nu^2)]$ is the effective bending modulus, $E$ is the Young's modulus, $\nu$ is the Poisson's ratio, $t_{\mathrm{eff, \ bend}}$ is the effective thickness of ribbons in bending, and $L_B$ is the length of each bending hinge. In our simulation, we use $C_B=10$ which is chosen such that our system matches the experimental rest shapes, and we use $E=3.1 \ \mathrm{GPa}$ and $\nu=0.38$ obtained through a tensile test of the Mylar\textsuperscript{\textregistered} sheets. For bending hinges of type I, the effective thickness is $t_{\mathrm{eff,  \ bend,  \ I}} = t$ (where $t=0.1905 \ \mathrm{mm}$ is the thickness of individual ribbons) because only one ribbon is bent in the longitudinal direction where orthogonal ribbons meet; while for bending hinges of type II, the effective thickness is $t_{\mathrm{eff, \ bend, \ II}} = \sqrt[3]{2}t$ because the effective bending modulus is a linear superposition of the individual moduli of the non-rigidly bonded woven ribbons \cite{Hibbeler2016Mechanics,peng2024analytic}. The stretching stiffness $K_A$ is given by 
\begin{equation}\label{eqn:KA}
K_A =C_A Y_{\mathrm{eff}} \frac{{S}}{CL_A(1 - \nu )}{\left( {\frac{h}{L_A}} \right)^{1/3}}= C_A Et_{\mathrm{eff,\ axial}} \frac{{S}}{CL_A(1 - \nu )}{\left( {\frac{h}{L_A}} \right)^{1/3}},
\end{equation}
where $C_A$ is a dimensionless constant for stretching stiffness calibration \cite{wo2023stiffening}, $Y_{\mathrm{eff}}=Et_{\mathrm{eff,\ axial}}$ is the effective stretching modulus, $t_{\mathrm{eff, \ axial}}$ is the effective thickness of ribbons in axial stretching, $L_A$ is the length of each bar, $S$ and $C$ are the area and perimeter, respectively, of the triangular panel associated with the current bar of interest, and $h$ is the height of the triangular panel. If a bar is shared by more than one triangle, its stretching stiffness is calculated by summing the stiffness contributed by each adjacent panel. In our simulation, we use $C_A=0.36$ which is chosen such that our system matches the experimental linear stiffnesses \cite{tu2024origami}. For bars in stretching, the effective thickness $t_{\mathrm{eff,\ axial}}=2t$ because without buckling or out-of-plane deformation, the woven ribbons and their continuous shell counterpart are assumed to exhibit the same axial behavior \cite{tu2025basketweaving}. 

Using the bar \& hinge mesh, we obtain the rest shapes of woven surfaces using a \textit{self-morphing} technique \cite{tu2025basketweaving}. We first model the initial geometry of a woven surface using its \textit{trivial shape} with the same topology. For example, we model the initial geometry of the snagged woven surface in Fig. \ref{fig:Model}(a) using a trivial non-smooth surface that consists of a rectangular substrate and two strip-like snags (Fig. \ref{fig:Model}(b)). We then set the stress-free angles of all bending hinges $\theta_{B}^{\mathrm{SF}}$ to be all $\pi$ ($180^{\circ}$), which is a flat rest state in contrast to the initial trivial shape where some ribbons are bent to $\pi/2$ ($90^{\circ}$). By changing the rest angles, we let the system address its own imbalance and eventually find the actual smooth rest shape that makes the resultant internal force zero. We use the incremental Newton-Raphson iterations to track the nonlinear equilibrium path of the system where we incrementally increase the stress-free angles from their initial values to $\pi$, and the woven surface gradually morphs from the trivial shape to its rest shape (Fig. \ref{fig:Model}(c)). The total potential energy $\Pi$ of a bar \& hinge system that includes $M$ hinges and $N$ bars can be computed as:
\begin{equation}\label{eqn:PotentialEnergy}
\begin{array}{l}
\begin{split}
\Pi &= U_{\mathrm{bending}} + U_{\mathrm{stretching}} + V_{\mathrm{external}}\\
 &= \sum\limits_{m = 1}^M {\frac{1}{2}{K_{B,m}}{{\left( {{\theta _{B,m}} - \theta_{B,m}^{\mathrm{SF}}} \right)}^2}}  + \sum\limits_{n = 1}^N {\frac{1}{2}{K_{A,n}}{{\left( {{L_{A,n}} - L_{A,n}^{\mathrm{SF}}} \right)}^2}} + {{\bf{f}}^T}{\bf{u}},
\end{split}
\end{array}
\end{equation}
where $U_{\mathrm{bending}}$ and $U_{\mathrm{stretching}}$ represent the internal bending energy and stretching energy computed based on the rotation of the bending hinges and deformation of the bars; while $V_{\mathrm{external}}$ represents the potential of the external load. The two quantities $\theta_{B,m}$ and $L_{A,n}$ represent the current hinge angles and bar lengths, while $\theta_{B}^{\mathrm{SF}}$ and $L_{A}^{\mathrm{SF}}$ represent the stress-free states of the same elements. In the incremental Newton-Raphson solver for self-morphing, we incrementally increase $\theta_{B}^{\mathrm{SF}}$, and we set $L_{A}^{\mathrm{SF}}$ to the initially calculated bar lengths based on the trivial shape. The two vectors $\bf{f}$ and $\bf{u}$ are the nodal load and nodal displacement vectors. In the self-morphing process, there is no external load so the $V_{\mathrm{external}}$ in Eq.~\ref{eqn:PotentialEnergy} is zero. Before we run the self-morphing solver, we set up boundary conditions which eliminate the six rigid body motions of the surface but do not impede any elastic deformations. Figure \ref{fig:Model}(b) shows the self-morphing process from the trivial to rest shape of the snagged woven sheet in Fig. \ref{fig:Model}(a).

\begin{figure}[!htb] 
\centering
\makebox[0pt]{\includegraphics[scale=1]{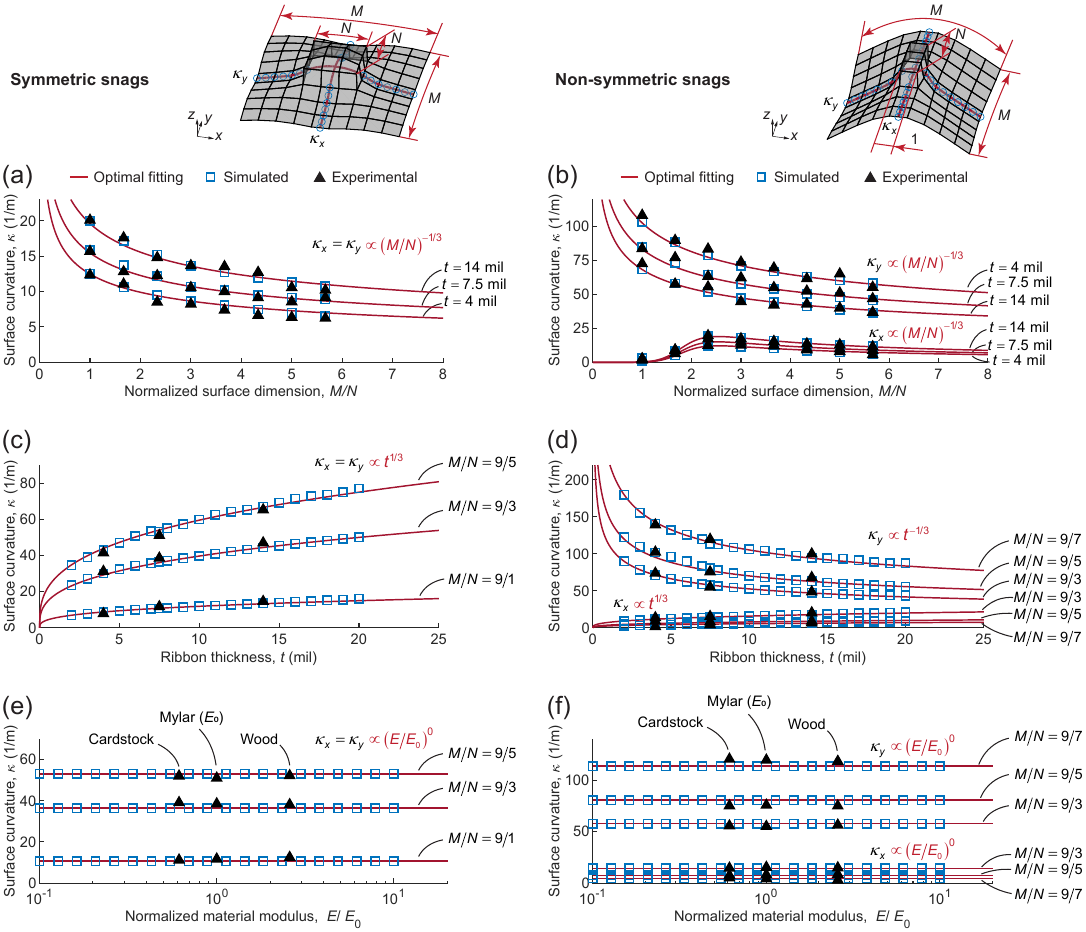}}
\caption{\textbf{Scaling relationships between the snag-enabled curvature and essential physical quantities of a woven system}. (a, b) shows the relationship between the surface curvature $\kappa_x$, $\kappa_y$ and the normalized surface dimension $M/N$; (c, d) shows the relationship between the surface curvature $\kappa_x$, $\kappa_y$ and the thickness of the ribbons $t$; (e, f) shows the relationship between the surface curvature $\kappa_x$, $\kappa_y$ and the normalized Young's modulus of the material $E/E_0$ where $E_0$ represents the Young's modulus of Mylar\textsuperscript{\textregistered}. The left column (a, c, e) shows the cases where the snag is rotationally symmetric (a square on the center), while the right column (b, d, f) shows the cases where the snag is rotationally non-symmetric (a rectangular strip on the center aligned with the \textit{y} axis where the width of the strip is always 1 unit and the length is $N$ units ($N=$ 3, 5, 7, ...)).}\label{fig:Scaling}
\end{figure}

\section{Scaling relationships}\label{Sec:Scaling}
Using the bar \& hinge model introduced in Section \ref{Sec:Modeling} and surface scanning, we numerically and experimentally study the scaling relationships between the geometry of snagged woven surfaces and essential design parameters. Our study is based on the two most typical snagged woven surfaces---one has a square substrate with a square snag at its center, which causes the sheet to bend symmetrically about both axes of symmetry (Fig. \ref{fig:Scaling}, left upper panel); while the other has a square substrate of the same dimensions but with a rectangular strip-like snag at the center, which causes the sheet to bend more about the \textit{y} axis (Fig. \ref{fig:Scaling}, right upper panel). We denote the design parameters including the surface dimension as $M$, snag dimension as $N$, ribbon thickness as $t$, material modulus as $E$; and we denote the geometric parameters including the maximum absolute curvature value of the substrate bending about the \textit{y} axis as $\kappa_y$ and about the $x$ axis as $\kappa_x$. The two curvatures $\kappa_y$ and $\kappa_x$ are calculated based on the coordinates of the sampled nodes on the sheet substrates and not on the snags, which are indicated by the blue circles shown on the top of Fig. \ref{fig:Scaling}. When we study the scaling relationship regarding a certain design parameter, all other design parameters remain constants unless stated otherwise. 

We first vary the surface dimension $M$ while keeping the snag dimension $N$ constant to reveal how the snag distributes the geometric frustration across surfaces of different length scales. For the symmetrically snagged surface (Fig. \ref{fig:Scaling}(a)), when the normalized surface dimension $M/N$ approaches infinity, the surface curvature approaches zero at a rate of $(M/N)^{-1/3}$. This slow decay suggests that the local influence of the snag persists even in larger sheets. For the non-symmetrically snagged surface (Fig. \ref{fig:Scaling}(b)), the bending curvature about the flexible axis $\kappa_y$ is always larger than the bending curvature about the stiff axis $\kappa_x$. A slight increase in the curvature $\kappa_x$ occurs at first because when the normalized surface dimension $M/N$ is small, the snag almost expands over the entire width along the $y$ axis, and in this case a larger surface relaxes the strong constraint imposed by the rectangular snag which leads to an increase in $\kappa_x$. However, eventually both $\kappa_x$ and $\kappa_y$ reduce with $(M/N)^{-1/3}$, and similar to the case of the symmetrically snagged surface the curvature persists for large sheets.

Next, we vary the ribbon thickness $t$ to reveal how thicker or thinner material affects the shapes of snagged woven surfaces. For the symmetrically snagged surface, both curvatures $\kappa_x$ and $\kappa_y$ increase with the ribbon thickness $t$ (Fig. \ref{fig:Scaling}(c)) because the self-restoring force (a bending dominated behavior) of the woven surface increases as we use thicker material. However, for the non-symmetrically snagged surface, the bending curvature about the stiff axis $\kappa_x$ increases with the ribbon thickness $t$ while the bending curvature about the flexible axis $\kappa_y$ decreases (Fig. \ref{fig:Scaling}(d)). The opposite trends of $\kappa_x$ and $\kappa_y$ occur as the thickness $t$ approaches zero---in this case the self-restoring force approaches zero, and thus the woven surface conforms to the snag and forms a kink where $\kappa_y$ approaches infinity and $\kappa_x$ approaches zero. Despite the difference caused by the symmetry of the snag, the surface curvatures of both surfaces change at a rate of $t^{1/3}$or $t^{-1/3}$. 

Finally, we vary the Young's modulus of the material $E$ to study how the shapes of snagged woven surfaces change when we use different types of materials. As Fig. \ref{fig:Scaling}(e) and (f) show, the surface curvature does not change with the Young's modulus $E$. For a structural system with a linear elastic material, the Young's modulus $E$ factors out of the energy functional as a simple multiplicative constant (see Eqs. \ref{eqn:KB}, \ref{eqn:KA}, and \ref{eqn:PotentialEnergy}). Therefore, the Young's modulus $E$ affects the internal energy of the system but not the rest shape. 

From the curve sets in Fig. \ref{fig:Scaling}(c) and (e), we also observe that a larger snag (a smaller $M/N$) leads to a larger curvature for the symmetrically snagged surface. For the non-symmetrically snagged surface (Fig. \ref{fig:Scaling}(d) and (f)), a larger snag makes the bending curvature about the flexible axis $\kappa_y$ larger while about the stiff axis $\kappa_x$ smaller.

\section{Inverse design and applications}\label{Sec:InverseDesign}
\begin{figure}[!htb] 
\centering
\includegraphics[width=17.8cm]{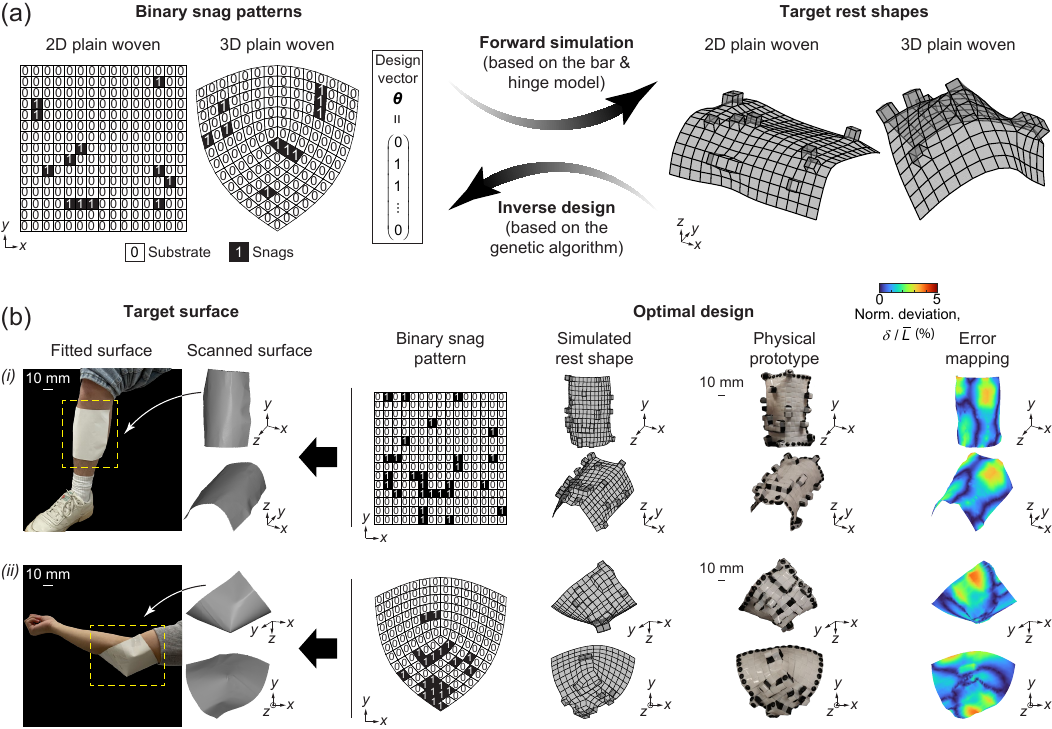}
\caption{\textbf{Inverse design of snagged woven surfaces}. (a) Schematic of the two-way process between the forward simulation of binary snag patterns and inverse design of target rest shapes; (b) Two examples of inverse design including (\textit{i}) an initially square sheet that deforms to fit a leg and (\textit{ii}) an initially conical surface that deforms to fit an elbow. The left of (b) shows the target surfaces including the photographs of fitted surfaces and the scanned surfaces, the right of (b) shows the optimal designs including the binary snag patterns, simulated rest shapes, photographs of physical prototypes, and error mappings (where the colormap represents the normalized deviation between the simulated rest shapes and scanned surfaces of the physical prototypes $\delta/\overline{L}$ where $\delta$ is the deviation (Hausdorff distances at each node) and $\overline{L}$ is the average length of all ribbons used; the simulated rest shapes are smoothed with snags removed from the mesh (see \textcolor{blue}{Fig. S1, Supplementary Information} for details)).}\label{fig:InverseDesign}
\end{figure}

We can make the snag patterns in Figs. \ref{fig:2DSnag} and \ref{fig:3DSnag} binary where we assign `0' to the substrate units and `1' to the snag units (Fig. \ref{fig:InverseDesign}, left panel). In this way, a snag pattern can be represented by a binary design vector $\boldsymbol{\theta} = {\left( {{\theta _1},{\theta _2}, \ldots ,{\theta _P}} \right)^T}$ where each element is 0 or 1 and the length $P$ is the total number of all square units in the original plain weave. Once a design vector is specified, we can obtain the geometry of the corresponding snagged woven surface based on our bar \& hinge model. This strategy also works the other way---we can inversely find an optimal design vector once a target surface is given. 

Here, we use the genetic algorithm \cite{goldberg1991comparative} to search the optimal design vector and thus to determine the optimal snag pattern. In this work, the size of the population in the genetic evolution is set to 200 and the maximum number of generations is set to 100; the initial population is constructed by an all-zero design vector and its 99 variants where in each variant a random number of elements are set to one; the number of elite individuals in each generation is set to 10 which is 5\% of the population size; the percentage of the population that undergoes crossover in each generation is set to 80\%; the probability of mutation is set to 0.3; the tolerance of the relative error for convergence is set to $10^{-6}$.  The fitness function $\mathcal{L}$ is given as
\begin{equation}\label{eqn:FIT}
\begin{array}{l}
\mathcal{L}(\boldsymbol{\theta} ;{\bf{S}}) = \underbrace {\frac{1}{Q}\sum\limits_{q = 1}^Q {{{\left| {{d_q}({\hat {\bf{S}}})} \right|}^2}} }_{{\rm{Deviation}}} + \lambda \underbrace {\sum\limits_{p = 1}^P {{\theta _p}} }_{{\rm{Sparsity}}}\\
{\rm{with }} \ {\bf{S}} = \left( {{\bf{x}},{\bf{y}},{\bf{z}}} \right) = F(\boldsymbol{\theta}).
\end{array}
\end{equation}

Our fitness function $\mathcal{L}$ (Eq. \ref{eqn:FIT}) has two terms. The first term `Deviation' represents how much the design surface deviates from the target surface. The $\boldsymbol{\theta }$ is the binary design vector with $\theta_p$ being the $p$-th element, $\bf{S}$ is the smoothed simulated snagged surface which is represented by a $Q\times3$ nodal coordinate matrix $ ({{\bf{x}},{\bf{y}},{\bf{z}}})$ and also a function of the design vector $F({\boldsymbol{\theta }})$; the $Q$ is the number of nodes compared on the simulated surface $\bf{S}$. The $\hat {\bf{S}}$ is the target surface, ${d_q}({\hat {\bf{S}}})$ is the Hausdorff distance \cite{huttenlocher1993comparing} from the $q$-th node in the simulated surface $\bf{S}$ to the target surface $\hat {\bf{S}}$. During each fitness evaluation, after we obtain a simulated snagged surface using our bar \& hinge model, we smooth the surface by removing the snag features from the surface mesh so that the deviation caused by snags will not affect the optimization (see \textcolor{blue}{Fig. S1, Supplementary Information} for details). The second term of the fitness function `Sparsity' represents how sparsely the snags are distributed over the woven surface. We combine the `Deviation' and `Sparsity' terms using the weight factor $\lambda$ in Eq. \ref{eqn:FIT}. A large $\lambda$ (1 or larger) leads to a surface with very few snags but the surface will not match the target very well. In this work, we set $\lambda$ to 0.3 to achieve an optimal inverse design using minimal snags so that we simplify our fabrication without sacrificing the accuracy.

Next, we give two real-life examples for the inverse design of woven exoskeleton surfaces (Fig. \ref{fig:InverseDesign}(b)). We use two flexible sticky paper sheets that initially have a 2D square shape and a 3D conical shape to fit onto the lower part of a human leg (Fig. \ref{fig:InverseDesign}(b, \textit{i}), left panels) and a human elbow (Fig. \ref{fig:InverseDesign}(b, \textit{ii}), left panels). Then, we scan these two surfaces and use the scans as target surfaces (which is $\hat {\bf{S}}$ in Eq. \ref{eqn:FIT}). The evolutionary search based on the genetic algorithm converges (see \textcolor{blue}{Fig. S2, Supplementary Information} for convergence plots), and our final optimal designs have an excellent match with the two target surfaces numerically and experimentally (Fig. \ref{fig:InverseDesign}(b), right panels). 

These inversely designed woven surfaces provide an excellent axial support stiffness and efficiently absorb energy under impact through the protruding snags \cite{tu2025basketweaving}. The woven nature of these snagged surfaces makes it easy to combine them with soft knitted fabrics. Therefore, our strategy for inverse design of snagged plain-woven surfaces can be easily applied to make customized human exoskeleton suits \cite{molinaro2024task,tian2023implant}. Future work could further make these exoskeleton surfaces `smart' and responsive to external environment by integrating active material in weaving. Besides the centimeter-scale applications shown here, our inverse design framework and the concepts of snags could be used for meter-scale structures in architectural systems \cite{bechthold2017materials} or micro-scale micro-electromechanical systems \cite{lin2024triboelectric,zhang2023elastocapillary}. 

\section{Conclusions}\label{Sec:Conclusion}
When snags in plain weaves are not accidental but designed on purpose, they have the potential to engineer the mechanical and geometric properties of the entire structure. In this work, we intentionally pull ribbons/fibers out of the plane and use circumferential weaves to incorporate snags into plain weaves based on pre-designed snag patterns where the sizes, shapes, and positions of each snag are dictated. Snags break the local in-plane surface of the weave and allow the fabric to deform out of the plane---these local geometric incompatibilities propagate through the entire weave and create a global 3D geometry from an initially 2D flat structure. For initially 3D curved plain weaves, snags can still be integrated to distort and tune the shapes of 3D spatial structures. We can simulate the geometry of snagged plain-woven surfaces using a mechanics-based reduced-order bar \& hinge model. 

When we vary the size of a snagged woven surface $M$ while maintaining the size of the snag, we have $\kappa \propto M^{-1/3}$ suggesting that the local snag's influence persists significantly even in larger surfaces. When we vary the thickness of the woven ribbons $t$, we have $\kappa \propto t^{1/3}$ for symmetrically snagged woven surfaces; while for non-symmetrically snagged woven surfaces, we have $\kappa \propto t^{1/3}$ for the stiffer axis of bending and $\kappa \propto t^{-1/3}$ for the more flexible axis of bending. When we explore materials with different Young's moduli $E$, the geometry of a snagged woven surface does not change. 

Finally, we use our mechanics model and an evolutionary algorithm to create an inverse design framework where we obtain optimal snag patterns which deform the surface to match with a given target shape. We give examples where snagged plain-woven surfaces are used to approximate the shapes of a human leg and elbow. Combined with active and electronic materials, our strategy of engineering snags into plain weaves has the potential to enable the next-generation fabrication of wearable conformal sensors, morphable lightweight composites, soft robotic skins, and more.

\section*{Acknowledgements}
The authors acknowledge support from the Air Force Office of Scientific Research under award number FA9550-22-1-0321. The authors also acknowledge helpful discussions with Dr. Yi Zhu. The paper reflects the views and opinions of the authors, and not necessarily those of the funding entities. 

\bibliography{Main_Text_V3}

\end{document}